\documentclass[letterpaper]{article} % DO NOT CHANGE THIS
\usepackage[submission]{aaai25}  % DO NOT CHANGE THIS
\usepackage{times}  % DO NOT CHANGE THIS
\usepackage{helvet}  % DO NOT CHANGE THIS
\usepackage{courier}  % DO NOT CHANGE THIS
\usepackage[hyphens]{url}  % DO NOT CHANGE THIS
\usepackage{graphicx} % DO NOT CHANGE THIS
\usepackage{natbib}  % DO NOT CHANGE THIS AND DO NOT ADD ANY OPTIONS TO IT
\usepackage{caption} % DO NOT CHANGE THIS AND DO NOT ADD ANY OPTIONS TO IT
\usepackage{algorithm}
\usepackage{algorithmic}

\usepackage{xcolor}
\newcommand{\answerYes}[1]{\textcolor{blue}{#1}} 
\newcommand{\answerNo}[1]{\textcolor{teal}{#1}} 
\newcommand{\answerNA}[1]{\textcolor{gray}{#1}}

\usepackage{booktabs} % For table rules
\usepackage{array}    % For custom column formatting
\usepackage{caption}  % For captions

\usepackage{graphicx} % for including images
\usepackage{subcaption} % for subfigure support (modern package)
\usepackage{amsmath} % for mathematical formulas
\makeatletter
\def\showauthors@on{T}
\makeatother

\usepackage{newfloat}
\usepackage{listings}
\DeclareCaptionStyle{ruled}{labelfont=normalfont,labelsep=colon,strut=off} % DO NOT CHANGE THIS
\floatstyle{ruled}
\newfloat{listing}{tb}{lst}{}
\floatname{listing}{Listing}
\title{The Emergence of Causal Curiosity from Prior Causal Belief Networks}
\author{
    Zhuoyu Shi\textsuperscript{\rm 1,2}, 
    Xintong Jiang\textsuperscript{\rm 1,2}, 
    Bohan Jiang\textsuperscript{\rm 3}, 
    Fred Morstatter\textsuperscript{\rm 1,2}
}
\affiliations{
    \textsuperscript{\rm 1}Thomas Lord Department of Computer Science, University of Southern California, USA\\
    \textsuperscript{\rm 2}Information Sciences Institute, University of Southern California, USA\\
        \textsuperscript{\rm 3}School of Computing and Augmented Intelligence, Arizona State University, USA\\

    zhuoyush@usc.edu, xintongj@usc.edu, bjiang14@asu.edu, morstatt@usc.edu
}

\begin{document}

\maketitle

\begin{abstract}
Causal curiosity is foundational to human cognition. It is the desire to understand why events happen, what mechanisms underlie them, and how outcomes can be explained or anticipated. It motivates exploration, sustains attention, and fuels the search for new knowledge. However, despite consensus on the importance of causal curiosity, little is known about how causal curiosity arises from existing belief systems. In our work, we examine how causal curiosity emerges from prior knowledge structures. With Reddit data from 2020 to 2023, and leveraging language models to extract cause-and-effect relationship pairs and identify causal curiosity-driven questions, our findings reveal that causal curiosity is not random or independent, but largely rooted in prior knowledge structure. Particularly, \textit{positive} nouns are more likely to be the object of causal curiosity. Moreover, our findings indicate that concepts that serve more as \textit{causes} are more likely to appear in causal curiosity than those that serve more as effects. Lastly, our results also suggest that causal curiosity emerges from the \textit{central} of the prior belief network. These novel insights reveal that causal curiosity is not random but systematically grounded in prior knowledge structures, and suggest their implication to facilitate the human learning process by motivating people to actively explore and construct deeper understandings rather than passively receiving information. 
\end{abstract}

\section{Introduction}

Curiosity is a foundational drive of human cognition. It motivates exploration, sustains attention, and fuels the search for new knowledge \cite{berlyne1960conflict, loewenstein1994psychology, kidd2015psychology}. From childhood through adulthood, curiosity not only supports learning but also shapes how individuals interact with their environment, seek information, and generate explanations. Curiosity is a central mechanism through which people bridge the gap between the known and the unknown, guiding inquiry toward meaningful and rewarding discoveries.

Within this broad domain, causal curiosity refers to the desire to understand why events happen, what mechanisms underlie them, and how outcomes can be explained or anticipated. Research in developmental psychology shows that children's earliest ``why''-questions reflect a drive to acquire causal explanations rather than mere descriptions \cite{callanan1992preschoolers, schulz2007serious, gopnik2004theory}. Causal inquiry is selective, often directed at surprising or epistemically rich events \cite{schulz2004causal, schulz2007preschool}. Philosophical and cognitive work further underscores the elevated status of causal explanations: they are more satisfying, more useful for generalization, and more actionable than purely correlational accounts \cite{lombrozo2006structure, keil2006explanation}.

However, despite the overwhelming consensus on the importance of causal curiosity and causal reasoning, little is known about how causal curiosity arises from existing belief systems. Causal beliefs, structured mental models of how cause and effect are connected, are the structures through which people explain, predict, and intervene in the world \cite{sloman2005causal, woodward2005making}. No study has systematically investigated how curiosity interacts with these belief networks in the real world: whether curiosity emerges independently or whether it is anchored in the causal structures that people have already constructed. Understanding this relationship is crucial, as it speaks to how epistemic gaps are recognized, how questions are generated, and how collective knowledge evolves.

In our work, we investigate this by examining causal curiosity at scale through the lens of online discourse. Using Reddit data from 2020–2023, we construct subreddit-specific causal belief networks from posts and comments, and then analyze curiosity-driven ``why''-questions to understand their relationship with prior causal structures. Specifically, we investigate the following research questions:

\begin{itemize}
\item \textbf{RQ1} – Is causal curiosity grounded in the existing knowledge structures?
\item \textbf{RQ2} – How does sentiment differ in the expression of causal curiosity?
\item \textbf{RQ3} – Does causal curiosity target more on causes or effects within prior belief networks? 
\item \textbf{RQ4} – Where is causal curiosity located within causal networks?
\end{itemize}

Initially, we extract cause-and-effect relationship pairs from Reddit posts and comments between 2020 and 2022 using a fine-tuned language model, constructing subreddit-specific causal belief networks as mental models. We then identify causal curiosity in ``why'' post titles from 2023 using LLaMA-3.3-70B, a large language model classifier, and validated with human annotators. With these in place, we conduct four analyses. First, we examine lexical overlap to assess whether curiosity emerges from existing knowledge structures. Second, we analyze differences in sentiment in expressions of causal curiosity. Third, we evaluate network directionality to understand whether curiosity is directed more toward causes or effects. Fourth, we investigate where curiosity is located within prior belief networks.  Our results show that causal curiosity is systematically structured: it overwhelmingly draws on pre-existing concepts, disproportionately targets positively valenced nouns, orients toward causal antecedents, and clusters around structurally central ideas. Together, these findings reveal that causal curiosity is not random but emerges from prior knowledge structures.

\section{Related Work}

\subsection{Causal Curiosity}

Curiosity has long been studied as a fundamental cognitive drive, motivating information-seeking and learning \cite{berlyne1960conflict, loewenstein1994psychology, kidd2015psychology}. Within this broad domain, causal curiosity refers to the desire to understand why events occur, how mechanisms operate, and what factors explain observed outcomes. Developmental psychology shows that even very young children spontaneously generate “why”-questions and are motivated to obtain causal explanations rather than mere associations \cite{callanan1992preschoolers, schulz2007serious, gopnik2004theory}. Such inquiries are not random but often selective, directed toward events that are surprising, ambiguous, or epistemically rich \cite{schulz2004causal, schulz2007preschool}.

Philosophers and cognitive scientists have emphasized the centrality of causality in explanation and sensemaking. Explanations that identify underlying causal mechanisms are typically judged as more satisfying and more useful than those that merely describe correlations or outcomes \cite{lombrozo2006structure, keil2006explanation}. In decision-making contexts, curiosity about causal structure is instrumental, helping individuals to generalize knowledge and plan effective interventions \cite{griffiths2005structure, liquin2020functional}.

\subsection{Causal Belief and Mental Models}
Causal beliefs are the mental representations people construct about how events, agents, and variables are linked through cause and effect. These causal mental models allow individuals to move beyond description, supporting explanation of observed outcomes, prediction of future events, and the design of effective interventions \cite{sloman2005causal, woodward2005making}. Unlike simple associations, causal beliefs provide structured knowledge: they identify mechanisms, dependencies, and hierarchies that make complex phenomena intelligible and navigable.

Causal reasoning is therefore central to human cognition, shaping how people organize and interpret their experiences \cite{gopnik2012reconstructing}. As cognitive scaffolds, causal models help individuals connect disparate pieces of information into coherent knowledge structures, enabling both rapid sensemaking and long-term learning \cite{murphy1985role, sloman2005causal}. This structuring power extends beyond the individual. Cultural narratives and social discourse transmit shared causal understandings, guiding collective reasoning about all domains as varied as economics, social behavior, etc \cite{shi2024diffusion, norenzayan2005psychological}.

\section{Data Description}
To investigate the emergence of causal curiosity, we leverage a comprehensive dataset from Reddit\footnote{\url{https://academictorrents.com/details/56aa49f9653ba545f48df2e33679f014d2829c10}}, which contains every publicly available post and comment from June 2005 through December 2023. For the purposes of this study, we focus on the most active and influential communities by selecting the top 200 subreddits, as ranked in Reddit’s official Best of Reddit\footnote{\url{https://www.reddit.com/best/communities/1/?rdt=52192}} list. This captures a broad spectrum of discourse ranging from news and politics to science, entertainment, and everyday life.

To understand the emergence of causal curiosity, we restrict our analysis to the four most recent calendar years: January 2020 through December 2023. This window balances scale with temporal relevance, providing both sufficient data to model community-level causal structures and a focused period to examine how curiosity emerges. This filtered corpus provides a robust and large-scale foundation for modeling the causal structures embedded in community-level discourse and for analyzing patterns of expressed curiosity, as following: (i) subreddit-specific causal networks (2020–2022) and (ii) expressed causal curiosity in 2023, as reflected in ``why" question-asking posts. During this three year period (2020 - 2022), there are a total of 81,952,918 posts and 1,407,574,558 associated comments across the selected subreddits. During 2023, there are a total of 17,196,212 posts.

\section{Computational Methods}
To investigate the emergence of causal curiosity, we analyze Reddit data from 2020 to 2023. Specifically, we use posts (titles and selftexts) and comments from 2020 to 2022 to build mental models: networks of cause-and-effect relations expressed in text. We then examine post titles from 2023 to detect and quantify expressions of causal curiosity within these communities.

\subsection{Extracting Cause-and-Effect Pairs}
To extract causal relationships from textual content, we apply a state-of-the-art deep learning pipeline developed by Priniski et al.~\cite{priniski2023pipeline}, based on the RoBERTa model~\cite{liu2019roberta}. This transformer-based architecture has been fine-tuned to detect causal relations and corresponding spans (i.e., the "cause" and "effect" pairs) in natural language. Unlike binary classifiers, this model identifies explicit spans for both the cause and the effect within text, allowing fine-grained semantic representation. The model achieves strong performance metrics, with an F1 score of 0.874, precision of 0.883, and recall of 0.865~\cite{priniski2023pipeline}.

\begin{table*}[t]
\centering
\small
\begin{tabular}{p{0.40\textwidth} p{0.55\textwidth}} 

\toprule

\multicolumn{1}{c}{\textbf{Category}} & \multicolumn{1}{c}{\textbf{Subreddits}} \\

\midrule

General Discussion and Questions & r/AskReddit, r/NoStupidQuestions, r/explainlikeimfive, r/unpopularopinion \\
\addlinespace[2.5pt]

Humor and Memes & r/memes, r/funny, r/facepalm, r/mildlyinfuriating \\
\addlinespace[2.5pt]

\textbf{Advice and Personal Topics} & \textbf{r/relationship\_advice, r/dating} \\
\addlinespace[2.5pt]

\textbf{Gaming} & \textbf{r/gaming, r/leagueoflegends, r/Eldenring, r/DnD} \\
\addlinespace[2.5pt]

\textbf{Media and Entertainment} & \textbf{r/OnePiece, r/movies} \\
\addlinespace[2.5pt]

\textbf{Sports} & \textbf{r/nba, r/nfl, r/formula1, r/soccer} \\
\addlinespace[2.5pt]

\textbf{Finance and Investing} & \textbf{r/personalfinance, r/CryptoCurrency, r/wallstreetbets} \\
\addlinespace[2.5pt]

News and Current Events & r/news, r/politics \\
\addlinespace[2.5pt]

Regional & r/europe \\
\addlinespace[2.5pt]

\bottomrule
\end{tabular}
\caption{Subreddits that meet minimal thresholds for meaningful analysis (at least 500 posts expressing causal curiosity in 2023, more than 1,000 distinct noun tokens in those posts, and over 100,000 cause-and-effect pairs in titles, selftexts, and comments in 2022) grouped by content category. Subreddits whose content is overly broad, abstract, general and diffuse are excluded, and the final subreddits are in bold. }
\label{tab:subreddits_by_category}
\end{table*}

We apply this model to all Reddit post titles, selftexts, and comments in 2022 first. For each cause and effect pair, we extract their head noun using Spacy \cite{spacy2}, and build the directed, weighted network of noun-based causal relations. These form the basis of belief networks, which represent mental models.

\subsection{Detecting Causal Curiosity}
To identify expressions of causal curiosity, we focus on Reddit post titles from 2023 that contain the token “why.” Interrogatives beginning with “why” are widely recognized as canonical markers of causal inquiry: they explicitly request explanations framed in terms of causes, reasons, or mechanisms, rather than mere descriptions or factual details~\cite{pearl2018book, lombrozo2006structure, frazier2009preschoolers}. Developmental psychology shows that children’s spontaneous “why”-questions reflect an intrinsic drive for causal understanding~\cite{callanan1992preschoolers, norenzayan2000culture}, and work in information-seeking theory highlights “why”-questions as prototypical signals of epistemic curiosity~\cite{loewenstein1994psychology}.

This linguistic cue yields 732,752 candidate titles in our dataset. However, not all such titles express genuine curiosity. Some function rhetorically, serve promotional purposes, or narrate personal experiences (e.g., ``Why you need an air fryer”). To disambiguate these cases, we apply LLaMA-3.3-70B \cite{dubey2024llama}, a transformer-based large language model, as classifier, to distinguish true curiosity-seeking questions from other uses of “why.”

Some posts contain only a title, while others include both a title and a selftext. The model is prompted accordingly for both variants (see Appendix for full prompt design). Out of 582,309 title-only posts and 150,443 title-and-selftext posts, there are 535,506 (91.96\%) and 133,448 (88.70\%) are identified as true curiosity-seeking questions, respectively. For each identified why quesion expressing causal curiosity, we extract the noun with Spacy \cite{spacy2} within the title.

\subsection{Validation of Causal Curiosity Detection}
To evaluate the reliability of our LLaMA-based classification pipeline, we conducted a human annotation study using the Prolific platform. A total of 200 Reddit posts were randomly sampled: 100 with title-only content and 100 with both title and selftext.

Annotators were provided with five labeled examples for training and evaluated on six test cases. Only those who correctly labeled at least four out of six examples were retained. Each qualified annotator labeled 25 posts and received \$4 in compensation.

Each post was independently annotated by three distinct raters, who answered the following binary question: “Does the given text contain direct and neutral questions that clearly seek answers, opinions, or experiences from others?” Final labels were determined via majority vote to reduce annotator bias. Annotation instructions and examples are included in Figure~\ref{fig:anno1} and \ref{fig:anno2} in the Appendix.

The LLaMA model achieved a precision of 91\%  on title-only posts and 88\%  on title-and-selftext posts. To further evaluate consistency among annotators, we calculate the inter-annotator agreement using Fleiss’ Kappa. The overall agreement is 78.0\% for title-only posts and 79.3\% for title-and-selftext posts. These results indicate robust performance, sufficient for large-scale identification of causal curiosity across Reddit communities.

\subsection{Data Finalization}
After extracting cause–and-effect pairs from 2022 and identifying causal curiosity in 2023, there are 26 subreddits (see Table~\ref{tab:subreddits_by_category}) meet minimal thresholds for meaningful analysis: at least 500 posts expressing causal curiosity in 2023, more than 1,000 distinct noun words in those posts, and over 100,000 cause-and-effect pairs in titles, selftexts, and comments in 2022. To ensure that the resulting mental models are interpretable and community-specific, we exclude subreddits whose content is overly broad, abstract, general and diffuse (i.g., General Discussion and Questions, Humor and Memes, News and Current Events, and Regional categories). This yields 15 subreddits for final analysis.

For each of these subreddits, we further extract and incorporate cause-and-effect pairs from 2020 and 2021, combining them with those from 2022 to construct comprehensive three-year belief networks. These directed, weighted networks serve as the mental models. There are a total of 15,019,623 cause-and-effect noun pairs extracted.

\section{RQ1 – Causal Curiosity Emerge from Existing Knowledge Structures}

To examine whether causal curiosity is grounded in existing knowledge structures, we analyze lexical overlap between curiosity-driven discourse in 2023 and causal belief networks constructed from 2020–2022. Specifically, we compute the percentage of nouns appearing in causal curiosity titles that also occur in each subreddit’s causal network.

As shown in Table~\ref{tab:subreddit_overlap_why}, all 15 subreddits display high congruence. The mean overlap rate is 91.7\% (SD = 3.1\%), with values ranging from 84.8\% (r/dating) to 95.8\% (r/soccer). Domains such as r/soccer, r/NFL, and r/personalfinance exhibit the strongest alignment, while even more dynamic or speculative communities such as r/CryptoCurrency, r/wallstreetbets, and r/dating maintain overlap levels above 85\%.

These results suggest that causal curiosity is not random or independent, but emerges within the contours of prior causal discourse. People tend to ask causal curiosity-driven questions about entities and concepts already embedded in their belief structures. Causal curiosity therefore appears emerge from epistemic infrastructure: the shared mental models that both constrain and enable new causal inquiry.

\section{RQ2 - Sentiment Differences within Causal Curiosity}
To examine how sentiment shapes the emergence of causal curiosity, we analyze the relative frequency with which nouns of varying valence (positive, neutral, negative) appear in curiosity-driven posts across subreddits. We compare the proportion of nouns that are included in 2023 causal curiosty questions to those that are not, stratified by sentiment category.

\begin{figure*}[tbhp]
\centering
\includegraphics[width=.95\linewidth]{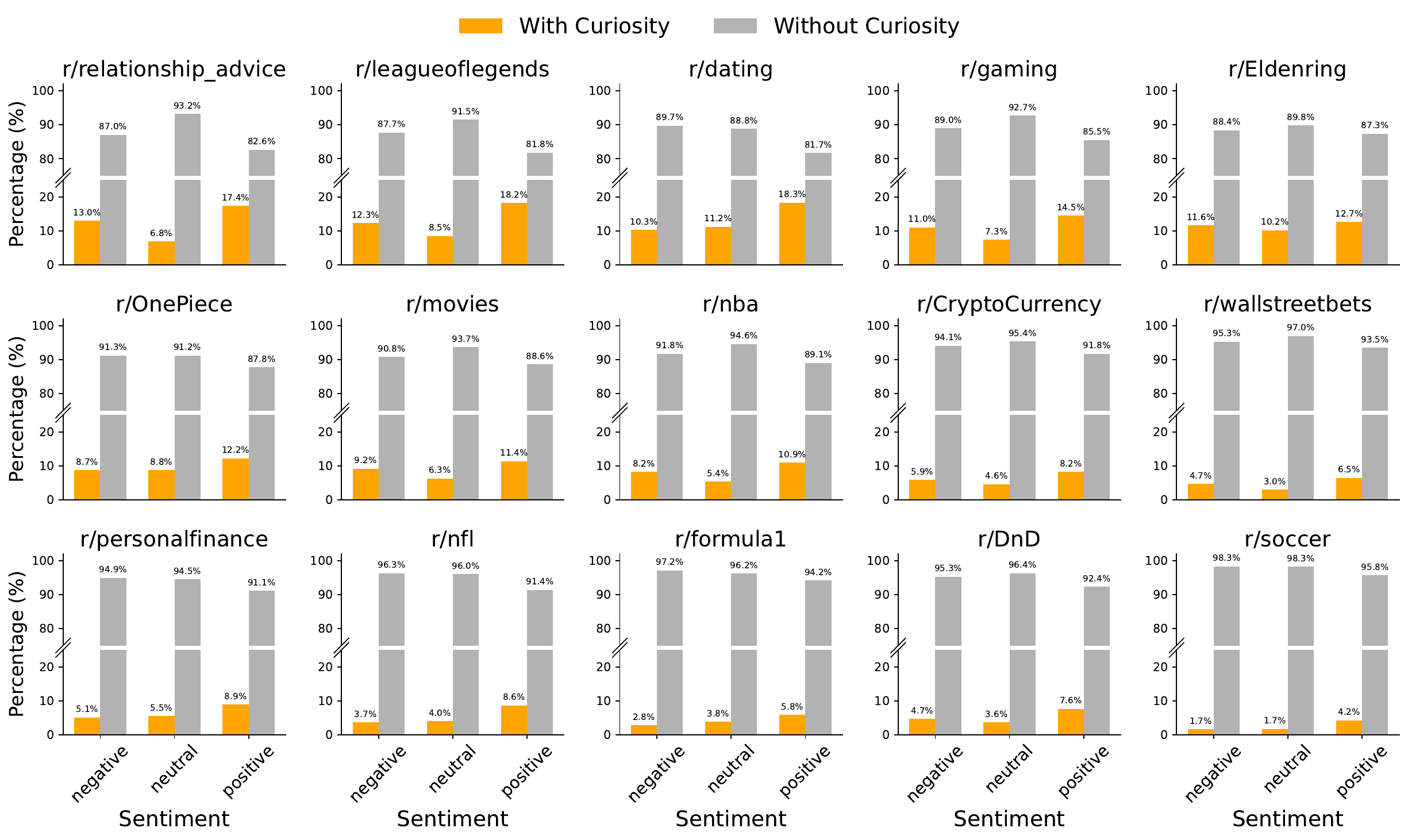}
\caption{Distribution of sentiment categories (negative, neutral, positive) in nouns appearing with and without causal curiosity across subreddits. Across all subreddits, positive nouns are disproportionately more likely to be the object of causal curiosity.}
\label{fig:RQ2}
\end{figure*}

As illustrated in Figure~\ref{fig:RQ2}, it suggests a consistent pattern across communities: positive nouns are disproportionately more likely to be the object of causal curiosity. For example, in dating, over 18\% of positive nouns appear in causal curiosoty questions, compared to only 10\% of negative nouns. This positivity-skew is especially prominent in socially or emotionally themed subreddits such as r/dating. More outcome-oriented or analytical communities, such as r/personalfinance, r/CryptoCurrency, and r/wallstreetbets, still show the trend, but with a narrower margin. Only a handful of subreddits  exhibit near-flat sentiment distributions, likely due to the external or performance-based nature of their discourse.

While one might intuitively expect people to be especially curious about negative events, e.g., to understand what went wrong, our results reveal the opposite: across all subreddits, positively valenced concepts are more likely to be targeted by causal curiosity than negative or neutral ones.

Curiosity is forward-looking and goal-oriented. Researchers ~\cite{loewenstein1994psychology, kidd2015psychology} conceptualize curiosity as often being instrumental: people ask questions not just to resolve confusion but to guide future behavior. From this perspective, positive outcomes are more worth understanding because they are replicable and desirable. Learning why something good happened can help one try to achieve similar results, e.g., ``Why did this relationship work?” or ``Why is this financial strategy effective?" By contrast, negative events may not offer the same actionability or emotional incentive for exploration. Some might already be explained (or over-explained), while others are avoided due to discomfort or perceived helplessness.

\begin{table}[t]
\centering
\small
\begin{tabular}
{>{\centering\arraybackslash}p{0.2\textwidth} >{\centering\arraybackslash}p{0.2\textwidth}} 

\toprule

\multicolumn{1}{c}{\centering Subreddit} & \multicolumn{1}{c}{\centering  Percentage (\%)} \\

\midrule

r/relationship\_advice & 91.92 \\
r/leagueoflegends & 88.74 \\
r/dating & 84.82 \\
r/gaming & 90.57 \\
r/Eldenring & 88.92 \\
r/OnePiece & 87.37 \\
r/movies & 91.66 \\
r/nba & 93.97 \\
r/CryptoCurrency & 92.58 \\
r/wallstreetbets & 90.96 \\
r/personalfinance & 94.22 \\
r/nfl & 95.10 \\
r/formula1 & 94.34 \\
r/DnD & 94.01 \\
r/soccer & 95.81 \\

\bottomrule
\end{tabular}
\caption{Percentage of nouns appearing in causal curiosity
titles in 2023 that also occur in each subreddit’s causal network in 2020-2022. All 15 subreddits show high congruence (M = 91.7\%, SD = 3.1\%).}
\label{tab:subreddit_overlap_why}
\end{table}

Research in cognitive and affective psychology has long documented a positivity bias in attention, memory, and evaluative judgment, particularly in everyday, non-threatening contexts. Individuals are more likely to process, remember, and engage with positive stimuli, especially when making sense of social and personally relevant information \cite{reed2014meta, cacioppo1999emotion}. The current finding that positive concepts more frequently serve as anchors for curiosity-driven questioning, aligns with this broader affective asymmetry in cognitive salience. This positivity bias also shapes what seems worth explaining: if an event is perceived as good, surprising, or exemplary, it may prompt more public ``why"-questions. In contrast, negative experiences may be framed as inevitable, random, or unworthy of elaboration. In this way, positively valenced concepts may appear more epistemically open to exploration, while negative ones are often narratively closed.

Curiosity also often serves an anticipatory function: people seek explanations to acquire actionable knowledge or to mentally simulate future possibilities \cite{kidd2015psychology, loewenstein1994psychology}. From this perspective, positive outcomes may elicit more curiosity because they are more likely to be seen as replicable, desirable, or strategic. That is, users may ask why something good happened not just to make sense of it, but to potentially reproduce or attain similar outcomes. This suggests that curiosity is not merely an epistemic impulse but also reflects instrumental rationality.

Causal curiosity may also be shaped by affective forecasting, the tendency to overestimate the emotional payoff of positive future outcomes \cite{wilson2003affective}. If individuals implicitly associate knowledge about positive entities with potential reward (e.g., success, satisfaction, improvement), then the act of asking ``why" becomes not only a cognitive but also an affective investment. This helps explain the observed pattern where positive nouns disproportionately anchor causal inquiries.

Together, these findings point to an important affective dimension in the structure of causal curiosity. Rather than arising evenly across conceptual space, causal curiosity is affectively selective, gravitating more strongly toward positively valenced concepts. This suggests that curiosity is not solely driven by knowledge gaps, but also by motivational landscapes, where the perceived value or emotional utility of an explanation shapes whether a question is even posed.

\section{RQ3 - Directionality in Causal Curiosity}
To understand the directional focus of causal curiosity, we examine whether individuals are more likely to ask ``why" about concepts that function as causes or as effects within subreddit-specific belief networks. We define this distinction operationally using causal graph topology: for each noun in the belief network, we compute its in-degree centrality (how often it appears as an effect) and out-degree centrality (how often it appears as a cause). We then calculate the difference score. 

For each node in the network, if in-degree $\geq$ out-degree, it implies that a concept is more often the recipient of causal links (i.e., an effect), whereas in-degree $<$ out-degree implies it is more often a cause in the network. This measure enables us to test whether curiosity gravitates more toward causal antecedents or consequences.

\begin{figure*}[tbhp]
\centering
\includegraphics[width=.88\linewidth]{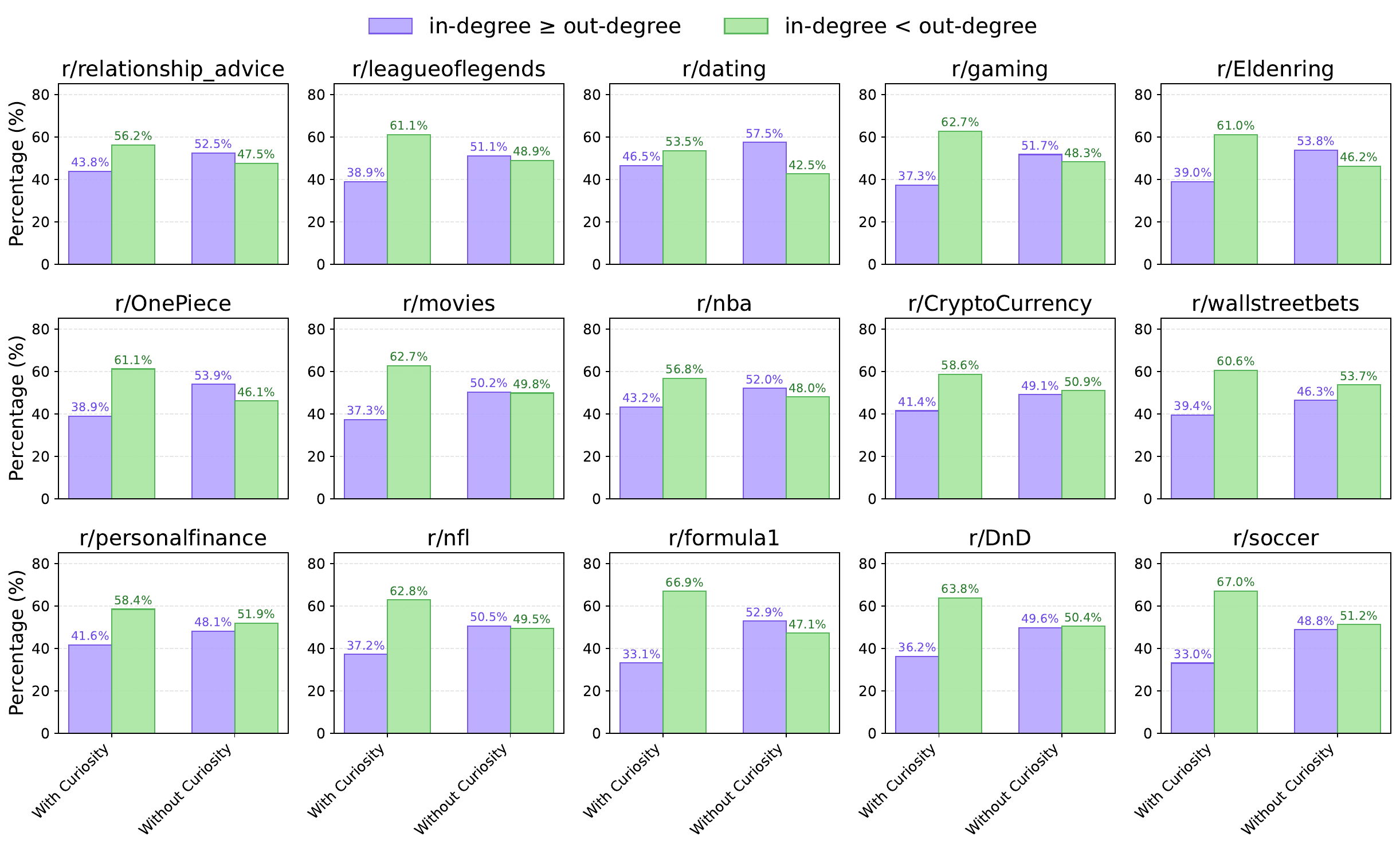}
\caption{The percentage distribution of nouns used in curiosity-driven titles versus non-curiosity cases, grouped by the directionality of their causal role (nouns functioning more as causes: in-degree $<$ out-degree, or functioning more as effects: in-degree $\geq$ out-degree). Concepts that serve more often as causes are significantly more likely to appear in causal curiosity-seeking questions.}
\label{fig:RQ3}
\end{figure*}

Figure~\ref{fig:RQ3} presents the percentage distribution of nouns used in curiosity-driven titles versus non-curiosity cases, grouped by the directionality of their causal role (in-degree $\geq$ out-degree, or in-degree $<$ out-degree).

Across all 15 subreddits, we observe a clear and robust trend in causal curiosity-seeking questions: nouns which in-degree $<$ out-degree is more than nouns in-degree $\geq$ out-degree. That is, concepts that serve more often as causes, are significantly more likely to appear in causal curiosity-seeking questions. For example, in formula1, 66.9\% of nouns used in curiosity questions are causal antecedents (out-degree dominant), compared to just 33.1\% for effect-like concepts. Similar trends appear in all subreddits, for example,  r/soccer (67.0\%),   r/DnD (63.8\%),  r/movies (62.7\%), etc.

This pattern suggests that causal curiosity is epistemically grounded in the search for generative mechanism. Rather than wondering about outcomes themselves (``why is X happening to me?"), users appear more often to ask about the forces, agents, or variables that produce those outcomes (``why does Y cause this?"). In graph-theoretic terms, curiosity is directed upstream, toward conceptual nodes that propagate influence within the community's causal mental model.

This resonates with cognitive theories of explanation that emphasize the centrality of causal primitives in sensemaking. Explanations that identify root causes are generally preferred, perceived as more satisfying, and more likely to be retained in memory \cite{lombrozo2006structure}. From a computational standpoint, this may reflect an optimization strategy: seeking to identify a small number of high-leverage causal factors that structure the wider semantic space.

The preference toward curiosity about causes may also reflect motivational and affective dynamics. Causes are often actionable: if one can understand what drives an outcome, one might be able to prevent, replicate, or influence it. Effects, by contrast, are often downstream observations: useful for evaluation, but less informative for intervention.

In this sense, causal curiosity reflects not only a desire for understanding, but a preference for explanatory control: a cognitive orientation toward concepts that offer leverage over the system.

This reveals that causal curiosity is systematically skewed toward concepts that serve as causes, rather than effects, in community discourse. This suggests  that curiosity is not only a response to informational gaps, but is strategically directed toward conceptual nodes that promise deeper explanatory or instrumental value.

\section{RQ4 - Structural Centrality and Causal Curiosity}
To examine how conceptual centrality influences causal curiosity, we analyse where do nouns in causal curiosity-driven questions appear in subreddit-level causal networks. We operationalize centrality in two ways: degree centrality  and betweenness centrality. For each measure, nouns are divided into 10 percentile bins, and we compute the proportion of curiosity-driven mentions in each bin.

To examine how conceptual centrality influences causal curiosity, we analyse where do nouns in causal curiosity-driven questions appear in subreddit-level causal networks. We operationalize centrality in two ways: betweenness centrality, which captures the extent to which a concept bridges distinct parts of the network, and degree centrality, which measures the overall number of direct causal connections. For each measure, nouns are ranked by centrality percentile and grouped into bins (0–10\%, 10-20\%, …, 90–100\%). We then compute the proportion of nouns in each bin that appear in 2023 curiosity-driven titles.

\begin{figure*}[tbhp]
\centering
\includegraphics[width=.99\linewidth]{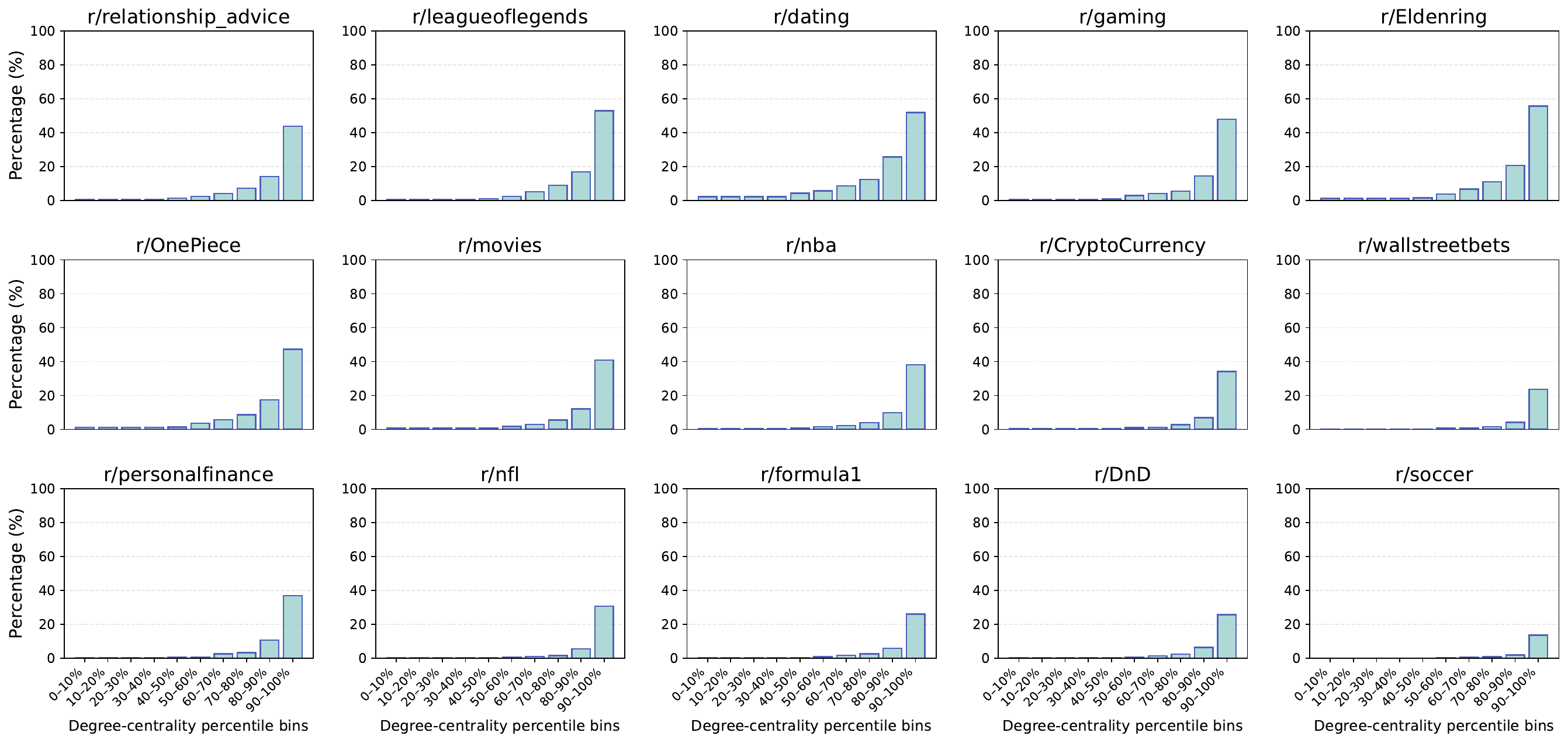}
\caption{Relationship between degree-centrality and causal curiosity. Nouns are binned by degree-centrality percentiles (0–10\%, 10–20\%, …, 90–100\%). The y-axis shows the percentage of nouns \textit{with curiosity} in each bin. Curiosity is strongly concentrated in the top 10\% of central concepts (M = 37.7\%, SD = 11.7\%).}
\label{fig:RQ4_degree}
\end{figure*}

\begin{figure*}[tbhp]
\centering
\includegraphics[width=.99\linewidth]{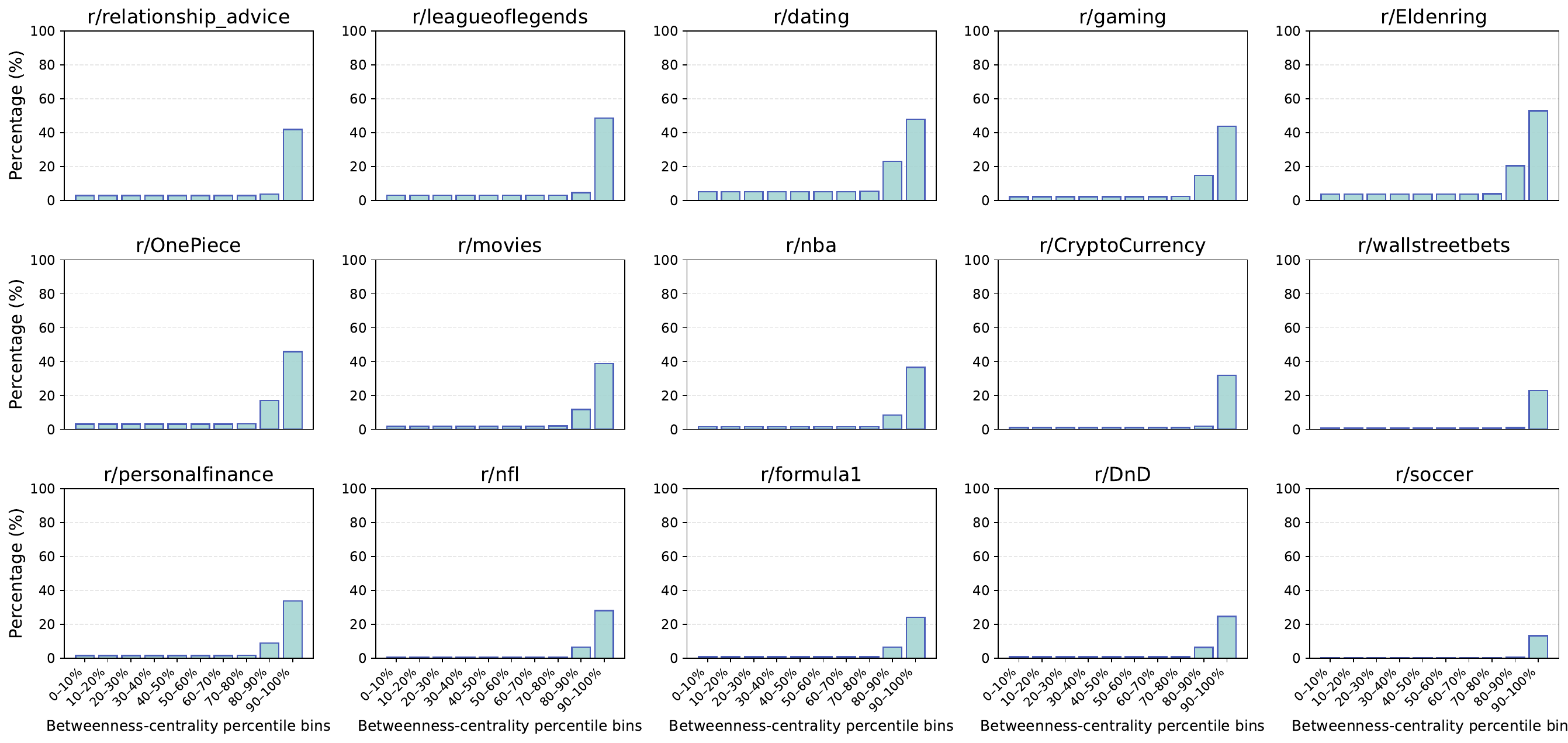}
\caption{ Relationship between betweenness-centrality and causal curiosity. Nouns are binned by betweenness-centrality percentiles (0–10\%, 10–20\%, …, 90–100\%). The y-axis shows the percentage of nouns \textit{with curiosity} in each bin. Curiosity disproportionately targets high-centrality nodes (top 10\%: M = 35.4\%, SD = 9.9\%), underscoring the role of bridging concepts in structuring explanation-seeking.}
\label{fig:RQ4_between}
\end{figure*}

Figures~\ref{fig:RQ4_degree} and~\ref{fig:RQ4_between} reveal a highly skewed distribution: curiosity is disproportionately concentrated on the most central concepts. On average, nouns in curiosity-driven questions account for 37.7\% in the top 10\% of degree centrality  (SD = 11.7\%), with subreddit-level values ranging from 13.7\% in r/soccer to 55.7\% in r/Eldenring. Betweenness centrality shows a nearly identical pattern: the top 10\% of nodes account for 35.4\% of curiosity-driven mentions (SD = 9.9\%), spanning from 13.2\% in r/soccer to 52.9\% in r/Eldenring.

Together, these results show that causal curiosity is scaffolded by structural centrality: users causal curiosity is primarily about concepts that are both semantically salient and structurally pivotal. Explanations are sought not for peripheral details, but for hubs and bridges that organize the community’s causal discourse.

These findings suggest that central concepts in the community’s causal model are more likely to attract curiosity. From a network-theoretic standpoint, betweenness centrality identifies concepts that function as bridges: those that lie on the shortest causal paths between other nodes. Their structural importance implies that they connect otherwise disjoint parts of the belief network.

This aligns with theories in cognitive science that frame curiosity as being directed toward epistemically pivotal or semantically rich entities: those that promise broader inferential payoff  \cite{stahl2015observing, nelson2014children}. Central concepts may serve as leverage points in mental models: by understanding them better, individuals can potentially reduce uncertainty across many conceptual areas.

Furthermore, central nodes may represent shared epistemic touchpoints: terms that are more frequently encountered, more publicly discussed, or more emotionally salient. As such, they may be more cognitively accessible and socially legitimate targets for question-asking behavior.

This suggests that causal curiosity is not randomly or uniformly distributed across concepts. Instead, it clusters around a small set of highly central nodes, those that structurally connect diverse parts of the causal network. This suggests that users are intuitively drawn to concepts that offer high epistemic value: central ideas that, once understood, may unlock broader causal insight.

\section{Discussion}

This study set out to investigate how causal curiosity emerges and unfolds within large-scale online discourse. By constructing subreddit-specific causal belief networks from 2020–2022 data and analyzing curiosity-driven ``why''-questions in 2023, we examined the interplay between community knowledge structures and the expression of curiosity. Four key findings provide insight into the structure, valence, directionality, and centrality of causal inquiry in digital communities.

% \subsection{Knowledge Structures Scaffold Curiosity}
Our first finding is that causal curiosity is overwhelmingly grounded in prior community knowledge structures. Nearly 92\% of nouns used in causal curiosity-driven questions were already embedded in subreddit causal networks from the previous three years. This indicates that causal curiosity does not arise randomly or in isolation. Instead, it develops within the framework of existing causal discussions. People tend to frame their curiosity-driven questions around concepts and entities already integrated into their belief systems. In this sense, causal curiosity appears to stem from epistemic infrastructure, the shared mental models that both limit and support new avenues of causal inquiry.

% \subsection{Sentiment Bias in Causal Curiosity}
Our second major finding is that causal curiosity is systematically skewed toward positive concepts. Across various domains, positive nouns were disproportionately likely to appear in causal curiosity-driven questions.

This result may initially appear counterintuitive. One might expect people to be especially curious about negative outcomes, seeking to understand failures or prevent future harms. Yet our findings align with broader psychological evidence showing a positivity bias in attention, memory, and evaluative judgment, particularly in everyday, non-threatening contexts \cite{reed2014meta, cacioppo1999emotion}. Positive outcomes are more actionable and motivating: asking why something worked offers a path toward replication, while negative outcomes may be dismissed as inevitable or avoided due to discomfort.

Our results also connect with theories of curiosity as anticipatory and instrumental. Curiosity often functions not just to resolve uncertainty but to guide future action \cite{kidd2015psychology}. If positive outcomes are perceived as replicable and rewarding, they become natural anchors for explanation-seeking. This suggests that causal curiosity is not simply about resolving gaps in understanding but about maximizing epistemic and affective value.

% \subsection{Directionality: Curiosity About Causes, Not Effects}

A third pattern concerns the directionality of curiosity. We found that causal curiosity-driven questions disproportionately target cause-like concepts (out-degree dominant nodes) rather than effect-like ones (in-degree dominant). Across all subreddits, curiosity clustered around antecedent concepts, the factors, agents, or mechanisms that produce outcomes, rather than the outcomes themselves.

This finding aligns with cognitive science research on explanation preferences. Explanations that identify causal antecedents are generally judged more satisfying, more useful, and more memorable than those focusing on effects \cite{lombrozo2006structure}. From a computational perspective, curiosity about causes may represent an optimization strategy: a small number of root causes can organize and explain a much larger set of effects. This makes curiosity about causes both epistemically efficient and practically valuable.

There is also a motivating dimension. Causes are more actionable: knowing what drives an outcome allows individuals to intervene, replicate, or control future events. Effects, by contrast, are descriptive observations: valuable for evaluation but less so for agency. In this sense, the directional focus of curiosity reflects not only a desire for understanding but also an implicit preference for explanatory control.

% \subsection{Centrality Bias in Causal Curiosity}

Our final finding highlights the structural dimension of curiosity. Both degree and betweenness centrality analyses revealed that curiosity is disproportionately concentrated on a small set of highly central concepts. On average, the most central 10\% of nodes accounted for more than one-third of all curiosity-driven mentions, while peripheral concepts were almost entirely absent.

This centrality bias underscores that curiosity is not randomly distributed across conceptual space but is guided by structural prominence. Central nodes function as hubs or bridges in the network: they connect otherwise disjoint parts of the belief system and thus offer high inferential leverage. Understanding them promises broader epistemic payoff, consistent with theories that curiosity is directed toward epistemically rich targets \cite{stahl2015observing, nelson2014children}.

The concentration of curiosity on central nodes also reflects the social dimension of explanation-seeking. Central concepts are more frequently discussed, more visible, and more cognitively accessible, making them natural focal points for public inquiry. In this sense, causal curiosity is both cognitively strategic and socially scaffolded: it gravitates toward nodes that are pivotal in organizing collective discourse.

% \subsection{Summary}
Taken together, these results paint a picture of causal curiosity as systematically structured rather than haphazard. Curiosity emerges from existing knowledge, is biased toward positive and rewarding concepts, is directed upstream toward causes rather than downstream toward effects, and clusters around central nodes that structure discourse.

\section{Limitations and Ethical Considerations}

The Reddit data\footnote{\url{https://academictorrents.com/details/56aa49f9653ba545f48df2e33679f014d2829c10}} used in this study may include personally identifiable information and posts containing offensive language. Additionally, biases could arise due to the platform’s specific user demographics \cite{olteanu2019social}. While our analysis avoids identifiable information and focuses on aggregate patterns, it is important to remain sensitive to issues of consent, privacy, and potential harms of recontextualizing community discourse. The dataset does not specify a license and is distributed through Academic Torrents, a U.S. 501(c)(3) non-profit organization that facilitates decentralized and reproducible research sharing. We obtained approval from our University’s Institutional Review Board (IRB) to conduct an annotation study with Prolific annotators. However, the outcomes of this work carry risks of misuse, for instance, in manipulation, political campaigns, and other contexts. Lastly, causal networks derived from text rely on automated extraction pipelines, which, despite strong performance, introduce noise through parsing errors, semantic ambiguity, or incomplete representations of belief structures.

\section{Future Work}
Future research can extend this work in several directions. First, refining linguistic models of curiosity detection to incorporate a broader array of question forms and pragmatic cues could provide a more comprehensive picture of epistemic inquiry. Similarly, multimodal extensions could examine how curiosity manifests through images, memes, or links.  

Second, longitudinal designs could trace how specific events (e.g., global crises, technological innovations, major sporting outcomes) reshape the landscape of causal curiosity, potentially altering which concepts become epistemically central.

\section{Conclusion}
This study suggests that causal curiosity emerges systematically within prior belief structures. By leveraging Reddit as a window into collective epistemic behavior, we show that curiosity is not evenly distributed but is strategically directed toward familiar, positive, causally generative, and central concepts. Curiosity emerges from existing knowledge, is biased toward positive and rewarding concepts, is directed upstream toward causes rather than downstream toward effects, and clusters around central nodes that structure discourse.

\bibliography{aaai25}

@article{cacioppo1999emotion,
  title={Emotion},
  author={Cacioppo, John T and Gardner, Wendi L},
  journal={Annual review of psychology},
  volume={50},
  number={1},
  pages={191--214},
  year={1999},
  publisher={Annual Reviews 4139 El Camino Way, PO Box 10139, Palo Alto, CA 94303-0139, USA}
}

@article{gopnik2012reconstructing,
  title={Reconstructing constructivism: causal models, Bayesian learning mechanisms, and the theory theory.},
  author={Gopnik, Alison and Wellman, Henry M},
  journal={Psychological bulletin},
  volume={138},
  number={6},
  pages={1085},
  year={2012},
  publisher={American Psychological Association}
}

@book{sloman2005causal,
  title={Causal models: How people think about the world and its alternatives},
  author={Sloman, Steven},
  year={2005},
  publisher={Oxford University Press}
}

@article{murphy1985role,
  title={The role of theories in conceptual coherence.},
  author={Murphy, Gregory L and Medin, Douglas L},
  journal={Psychological review},
  volume={92},
  number={3},
  pages={289},
  year={1985},
  publisher={American Psychological Association}
}

@article{norenzayan2005psychological,
  title={Psychological universals: What are they and how can we know?},
  author={Norenzayan, Ara and Heine, Steven J},
  journal={Psychological bulletin},
  volume={131},
  number={5},
  pages={763},
  year={2005},
  publisher={American Psychological Association}
}

@inproceedings{priniski2023pipeline,
  title={Pipeline for modeling causal beliefs from natural language},
  author={Priniski, John and Verma, Ishaan and Morstatter, Fred},
  booktitle={Association for Computational Linguistics (Volume 3: System Demonstrations)},
  pages={436--443},
  year={2023}
}

@article{liu2019roberta,
  title={Roberta: A robustly optimized bert pretraining approach},
  author={Liu, Yinhan and Ott, Myle and Goyal, Naman and Du, Jingfei and Joshi, Mandar and Chen, Danqi and Levy, Omer and Lewis, Mike and Zettlemoyer, Luke and Stoyanov, Veselin},
  journal={arXiv preprint arXiv:1907.11692},
  year={2019}
}

@article{olteanu2019social,
  title={Social data: Biases, methodological pitfalls, and ethical boundaries},
  author={Olteanu, Alexandra and Castillo, Carlos and Diaz, Fernando and K{\i}c{\i}man, Emre},
  journal={Frontiers in big data},
  volume={2},
  pages={13},
  year={2019},
  publisher={Frontiers Media SA}
}

@misc{fair,
    title="The FAIR Data principles",
year = 2020,
    author="{FORCE11}",
howpublished="\url{https://force11.org/info/the-fair-data-principles/}"
}

@article{gebru2021datasheets,
  title={Datasheets for datasets},
  author={Gebru, Timnit and Morgenstern, Jamie and Vecchione, Briana and Vaughan, Jennifer Wortman and Wallach, Hanna and Iii, Hal Daum{\'e} and Crawford, Kate},
  journal={Communications of the ACM},
  volume={64},
  number={12},
  pages={86--92},
  year={2021},
  publisher={ACM New York, NY, USA}
}

@article{loewenstein1994psychology,
  title={The psychology of curiosity: A review and reinterpretation.},
  author={Loewenstein, George},
  journal={Psychological bulletin},
  volume={116},
  number={1},
  pages={75},
  year={1994},
  publisher={American Psychological Association}
}

@article{kidd2015psychology,
  title={The psychology and neuroscience of curiosity},
  author={Kidd, Celeste and Hayden, Benjamin Y},
  journal={Neuron},
  volume={88},
  number={3},
  pages={449--460},
  year={2015},
  publisher={Elsevier}
}

@article{reed2014meta,
  title={Meta-analysis of the age-related positivity effect: age differences in preferences for positive over negative information.},
  author={Reed, Andrew E and Chan, Larry and Mikels, Joseph A},
  journal={Psychology and aging},
  volume={29},
  number={1},
  pages={1},
  year={2014},
  publisher={American Psychological Association}
}

@article{wilson2003affective,
  title={Affective forecasting},
  author={Wilson, Timothy D and Gilbert, Daniel T},
  journal={Advances in experimental social psychology},
  volume={35},
  number={35},
  pages={345--411},
  year={2003},
  publisher={Elsevier Academic Press}
}

@article{lombrozo2006structure,
  title={The structure and function of explanations},
  author={Lombrozo, Tania},
  journal={Trends in cognitive sciences},
  volume={10},
  number={10},
  pages={464--470},
  year={2006},
  publisher={Elsevier}
}

@article{stahl2015observing,
  title={Observing the unexpected enhances infants’ learning and exploration},
  author={Stahl, Aimee E and Feigenson, Lisa},
  journal={Science},
  volume={348},
  number={6230},
  pages={91--94},
  year={2015},
  publisher={American Association for the Advancement of Science}
}

@article{nelson2014children,
  title={Children’s sequential information search is sensitive to environmental probabilities},
  author={Nelson, Jonathan D and Divjak, Bojana and Gudmundsdottir, Gudny and Martignon, Laura F and Meder, Bj{\"o}rn},
  journal={Cognition},
  volume={130},
  number={1},
  pages={74--80},
  year={2014},
  publisher={Elsevier}
}

@article{frazier2009preschoolers,
  title={Preschoolers’ search for explanatory information within adult--child conversation},
  author={Frazier, Brandy N and Gelman, Susan A and Wellman, Henry M},
  journal={Child development},
  volume={80},
  number={6},
  pages={1592--1611},
  year={2009},
  publisher={Wiley Online Library}
}

@article{callanan1992preschoolers,
  title={Preschoolers' questions and parents' explanations: Causal thinking in everyday activity},
  author={Callanan, Maureen A and Oakes, Lisa M},
  journal={Cognitive development},
  volume={7},
  number={2},
  pages={213--233},
  year={1992},
  publisher={Elsevier}
}

@article{norenzayan2000culture,
  title={Culture and causal cognition},
  author={Norenzayan, Ara and Nisbett, Richard E},
  journal={Current directions in psychological science},
  volume={9},
  number={4},
  pages={132--135},
  year={2000},
  publisher={SAGE Publications Sage CA: Los Angeles, CA}
}

@book{pearl2018book,
  title={The book of why: the new science of cause and effect},
  author={Pearl, Judea and Mackenzie, Dana},
  year={2018},
  publisher={Basic books}
}

@book{woodward2005making,
  title={Making things happen: A theory of causal explanation},
  author={Woodward, James},
  year={2005},
  publisher={Oxford university press}
}

@article{berlyne1960conflict,
  title={Conflict, arousal, and curiosity.},
  author={Berlyne, Daniel E},
  year={1960},
  publisher={McGraw-Hill Book Company}
}

@article{schulz2007serious,
  title={Serious fun: preschoolers engage in more exploratory play when evidence is confounded.},
  author={Schulz, Laura E and Bonawitz, Elizabeth Baraff},
  journal={Developmental psychology},
  volume={43},
  number={4},
  pages={1045},
  year={2007},
  publisher={American Psychological Association}
}

@article{gopnik2004theory,
  title={A theory of causal learning in children: causal maps and Bayes nets.},
  author={Gopnik, Alison and Glymour, Clark and Sobel, David M and Schulz, Laura E and Kushnir, Tamar and Danks, David},
  journal={Psychological review},
  volume={111},
  number={1},
  pages={3},
  year={2004},
  publisher={American Psychological Association}
}

@article{schulz2004causal,
  title={Causal learning across domains.},
  author={Schulz, Laura E and Gopnik, Alison},
  journal={Developmental psychology},
  volume={40},
  number={2},
  pages={162},
  year={2004},
  publisher={American Psychological Association}
}

@article{schulz2007preschool,
  title={Preschool children learn about causal structure from conditional interventions},
  author={Schulz, Laura E and Gopnik, Alison and Glymour, Clark},
  journal={Developmental science},
  volume={10},
  number={3},
  pages={322--332},
  year={2007},
  publisher={Wiley Online Library}
}

@inproceedings{shi2024diffusion,
  title={The Diffusion of Causal Language in Social Networks},
  author={Shi, Zhuoyu and Morstatter, Fred},
  booktitle={Proceedings of the International AAAI Conference on Web and Social Media},
  volume={18},
  pages={1422--1435},
  year={2024}
}

@article{keil2006explanation,
  title={Explanation and understanding},
  author={Keil, Frank C},
  journal={Annu. Rev. Psychol.},
  volume={57},
  number={1},
  pages={227--254},
  year={2006},
  publisher={Annual Reviews}
}

@article{griffiths2005structure,
  title={Structure and strength in causal induction},
  author={Griffiths, Thomas L and Tenenbaum, Joshua B},
  journal={Cognitive psychology},
  volume={51},
  number={4},
  pages={334--384},
  year={2005},
  publisher={Elsevier}
}

@article{liquin2020functional,
  title={A functional approach to explanation-seeking curiosity},
  author={Liquin, Emily G and Lombrozo, Tania},
  journal={Cognitive Psychology},
  volume={119},
  pages={101276},
  year={2020},
  publisher={Elsevier}
}

@software{spacy2,
  author       = {Matthew Honnibal and
                  Ines Montani and
                  Sofie Van Landeghem and
                  Adriane Boyd},
  title        = {spaCy: Industrial-strength Natural Language Processing in Python},
  month        = nov,
  year         = 2020,
  publisher    = {Zenodo},
  doi          = {10.5281/zenodo.1212303},
  url          = {https://doi.org/10.5281/zenodo.1212303}
}

@article{dubey2024llama,
  title={The llama 3 herd of models},
  author={Dubey, Abhimanyu and Jauhri, Abhinav and Pandey, Abhinav and Kadian, Abhishek and Al-Dahle, Ahmad and Letman, Aiesha and Mathur, Akhil and Schelten, Alan and Yang, Amy and Fan, Angela and others},
  journal={arXiv e-prints},
  pages={arXiv--2407},
  year={2024}
}

\section{Checklist}

% \subsection{Paper Checklist to be included in your paper}

\begin{enumerate}

\item For most authors...
\begin{enumerate}
    \item  Would answering this research question advance science without violating social contracts, such as violating privacy norms, perpetuating unfair profiling, exacerbating the socio-economic divide, or implying disrespect to societies or cultures?
    \answerYes{Yes, please see the Limitation and Ethical Considerations.}
  \item Do your main claims in the abstract and introduction accurately reflect the paper's contributions and scope?
    \answerYes{Yes, please see the Abstract and the Introduction.}
   \item Do you clarify how the proposed methodological approach is appropriate for the claims made? 
    \answerYes{Yes, please see the paper.}
   \item Do you clarify what are possible artifacts in the data used, given population-specific distributions?
   \answerYes{Yes, please see the Limitations and Ethical Considerations.}
  \item Did you describe the limitations of your work?
    \answerYes{Yes, please see the Limitations and Ethical Considerations.}
  \item Did you discuss any potential negative societal impacts of your work?
    \answerYes{Yes, please see the Limitations and Ethical Considerations.}
      \item Did you discuss any potential misuse of your work?
    \answerYes{Yes, please see the Limitations and Ethical Considerations.}
    \item Did you describe steps taken to prevent or mitigate potential negative outcomes of the research, such as data and model documentation, data anonymization, responsible release, access control, and the reproducibility of findings?
    \answerYes{Yes, please see the Limitations and Ethical Considerations.}
  \item Have you read the ethics review guidelines and ensured that your paper conforms to them?
    \answerYes{Yes, our paper conforms to the ethics review guidelines.}
\end{enumerate}

\item Additionally, if your study involves hypotheses testing...
\begin{enumerate}
  \item Did you clearly state the assumptions underlying all theoretical results?
    \answerNA{NA}
  \item Have you provided justifications for all theoretical results?
    \answerNA{NA}
  \item Did you discuss competing hypotheses or theories that might challenge or complement your theoretical results?
    \answerNA{NA}
  \item Have you considered alternative mechanisms or explanations that might account for the same outcomes observed in your study?
    \answerNA{NA}
  \item Did you address potential biases or limitations in your theoretical framework?
    \answerNA{NA}
  \item Have you related your theoretical results to the existing literature in social science?
    \answerNA{NA}
  \item Did you discuss the implications of your theoretical results for policy, practice, or further research in the social science domain?
    \answerNA{NA}
\end{enumerate}

\item Additionally, if you are including theoretical proofs...
\begin{enumerate}
  \item Did you state the full set of assumptions of all theoretical results?
    \answerNA{NA}
	\item Did you include complete proofs of all theoretical results?
    \answerNA{NA}
\end{enumerate}

\item Additionally, if you ran machine learning experiments...
\begin{enumerate}
  \item Did you include the code, data, and instructions needed to reproduce the main experimental results (either in the supplemental material or as a URL)?
    \answerNA{NA}
  \item Did you specify all the training details (e.g., data splits, hyperparameters, how they were chosen)?
    \answerNA{NA}
     \item Did you report error bars (e.g., with respect to the random seed after running experiments multiple times)?
    \answerNA{NA}
	\item Did you include the total amount of compute and the type of resources used (e.g., type of GPUs, internal cluster, or cloud provider)?
    \answerNA{NA}
     \item Do you justify how the proposed evaluation is sufficient and appropriate to the claims made? 
    \answerNA{NA}
     \item Do you discuss what is ``the cost`` of misclassification and fault (in)tolerance?
    \answerNA{NA}
  
\end{enumerate}

\item Additionally, if you are using existing assets (e.g., code, data, models) or curating/releasing new assets, \textbf{without compromising anonymity}...
\begin{enumerate}
  \item If your work uses existing assets, did you cite the creators?
    \answerYes{Yes, we cited them, please see the Dataset Description.}
  \item Did you mention the license of the assets?
    \answerYes{Yes, please see the Limitations and Ethical Considerations. We did not mention the license for the models because they are publicly available.}
  \item Did you include any new assets in the supplemental material or as a URL?
    \answerNo{No, we don't have any.}
  \item Did you discuss whether and how consent was obtained from people whose data you're using/curating?
    \answerYes{Yes, please see the Limitations and Ethical Considerations.}
  \item Did you discuss whether the data you are using/curating contains personally identifiable information or offensive content?
    \answerYes{Yes, please see the Limitation and Ethical Considerations.}
\item If you are curating or releasing new datasets, did you discuss how you intend to make your datasets FAIR (see \citet{fair})?
\answerNA{NA}
\item If you are curating or releasing new datasets, did you create a Datasheet for the Dataset (see \citet{gebru2021datasheets})? 
\answerNA{NA}
\end{enumerate}

\item Additionally, if you used crowdsourcing or conducted research with human subjects, \textbf{without compromising anonymity}...
\begin{enumerate}
  \item Did you include the full text of instructions given to participants and screenshots?
    \answerYes{Yes, please see the Validation and the Appendix.}
  \item Did you describe any potential participant risks, with mentions of Institutional Review Board (IRB) approvals?
    \answerYes{Yes, we obtained IRB approval from our University for our annotation experiment. Please see the Limitations and Ethical Considerations.}
  \item Did you include the estimated hourly wage paid to participants and the total amount spent on participant compensation?
   \answerYes{Yes, please see the Appendix (Annotation).}
   \item Did you discuss how data is stored, shared, and deidentified?
   \answerYes{Yes, we discussed how data was stored, shared, and deidentified. Please see the Limitations and Ethical Considerations.}
\end{enumerate}

\end{enumerate}

\section{Appendix}
\subsection{a. Prompt}

\subsubsection{a1. Prompt for title only posts}
\begin{quote}
You are an AI trained to classify forum titles based on their type. There are two categories:

- A: Direct and neutral questions that clearly seek answers, opinions, or experiences from others. These are often in a form of question. The tone is conversational and inquisitive.

 - B: Titles that appear to be article headlines, personal statements, explanations, or opinions. These do not expect an answer and are usually declarative, such as `Why [someone] is [something]', `Now I know why...', `10 reasons why...' or `Why should/shouldn't/don't you...'.
 
Instructions:
For each title, only return ``A" or ``B"—no additional text, explanations, or formatting.
 
Classify the following input:

Title: `\{title\}'
 
Output:
Category (ONLY return `A' or `B' with NO extra text)
\end{quote}

\subsubsection{a2. Prompt for title-and-selftext post}
\begin{quote}
You are an AI assistant trained to classify forum posts based on their Title and Text. Your task is to determine which of the following categories the given Title and Text belong to:

- A: The Title is a question, and the Text provides additional context or explanation related to the question.

- B: The Title is a question, but the Text is actually an answer to that question. The Text is usually long.

Instructions:
Carefully analyze both the Title and Text. Only return ``A" or ``B", no additional text, explanations, or formatting.
 
Classify the following input:

Title: `\{title\}'

Text: `\{text\}'
 
Output:
 Category (ONLY return `A' or `B' with NO extra text)

\end{quote}

\subsection{b. Annotation}
We spent a total of \$170.67 on the annotation task, which included \$42.67 in service fees charged by Prolific. Each annotator was assigned 25 posts to annotate via a Google Forms link distributed through Prolific and received a payment of \$4. Prolific handled all participant recruitment and automatically de-identified participant data to ensure privacy.

\begin{figure}[tbhp]
\centering
\includegraphics[width=0.9\linewidth]{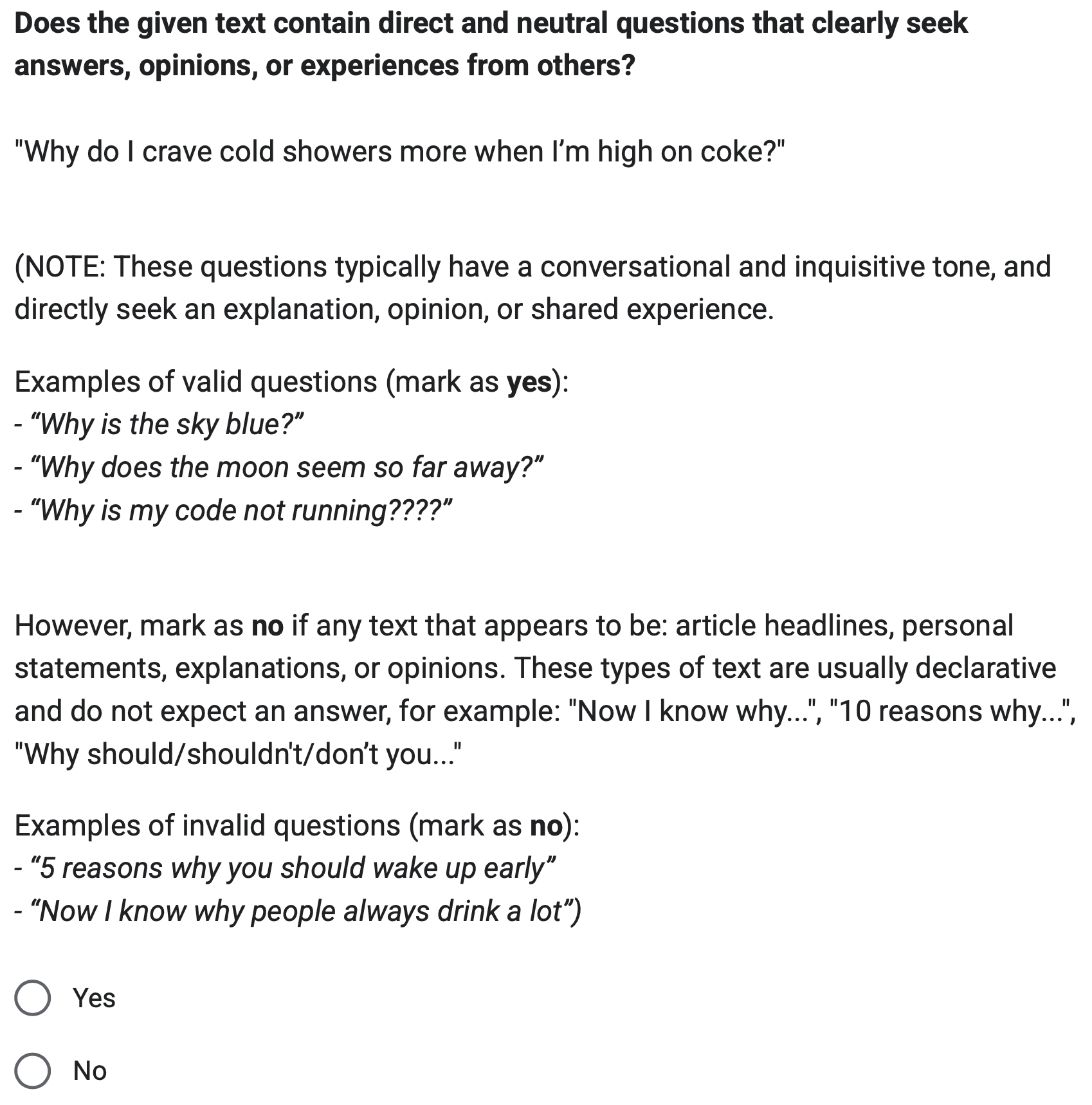}

\caption{Annotation example for title-only posts}
\label{fig:anno1}
\end{figure}

\begin{figure}[tbhp]
\centering
\includegraphics[width=0.9\linewidth]{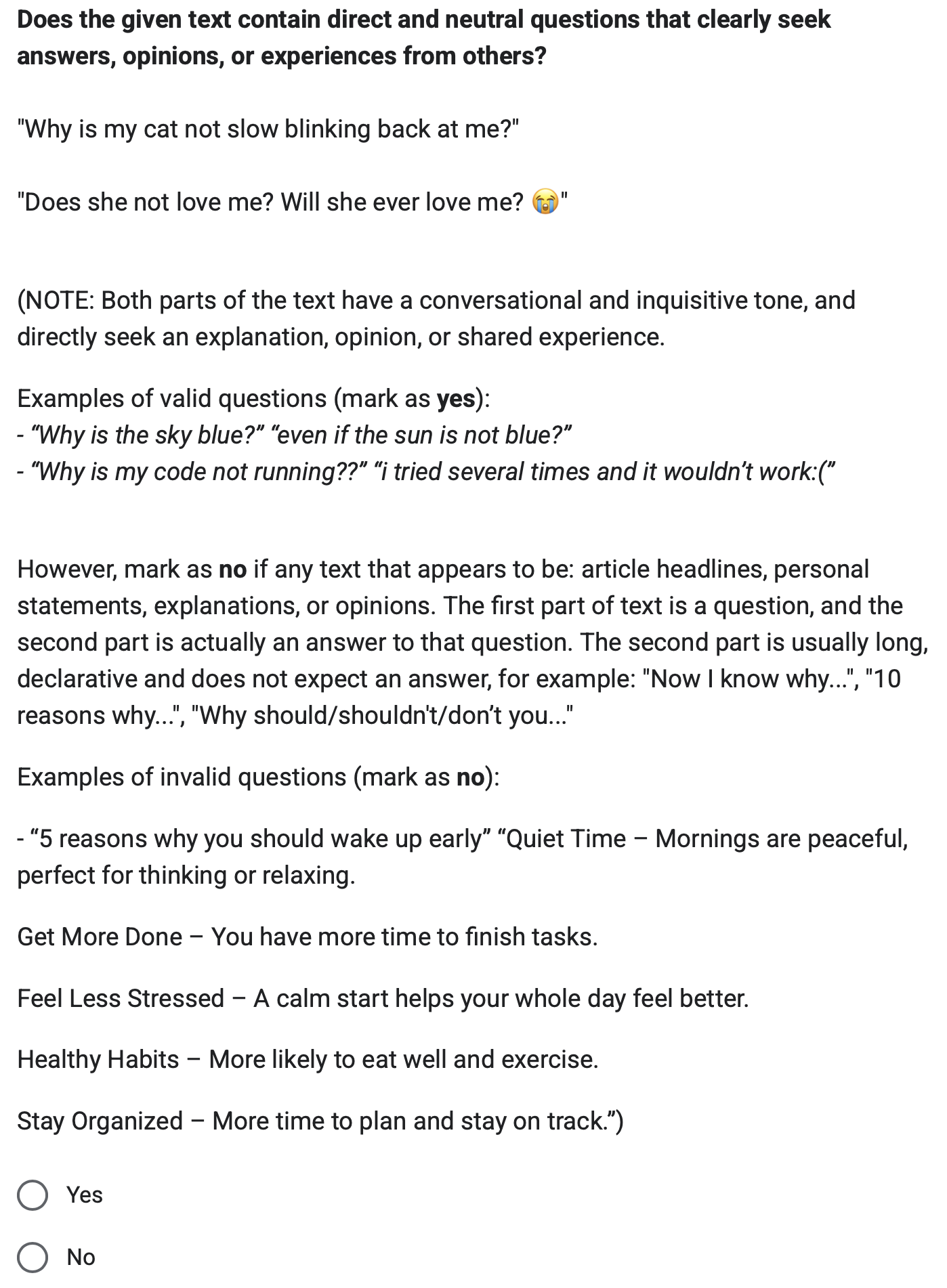}

\caption{Annotation example for title-and-selftext posts}
\label{fig:anno2}
\end{figure}

\end{document}